**A model for African trypanosome cell motility and quantitative description of flagellar dynamics**


**Nathan R. Hutchings[†*] and Andrei Ludu[‡*]**
Northwestern State University

[†] Department of Biological Sciences
[‡] Department of Chemistry and Physics
[*] Interdisciplinary Experimentation and Scholarship (IDEAS) program

Address correspondence to:

Dr. Nathan R. Hutchings
226B Bienvenu Hall
Northwestern State University
Natchitoches, Louisiana 71497
318.357.6514 phone
318.357.4518 fax
hutchingsn@nsula.edu

Key words:    Flagellum, Cell motility, Modeling, Non-linear, wave, waveform, Trypanosome, Trypanosoma, Crithidia, Sperm, Cilia, Swimming, Axoneme, Paraflagellar rod, Sleeping sickness, Trypanosomiasis, Physical, Geometry, Dynamics, conic, hyperbola, ellipse, Korteweg-de Vreis (mKdV)




Short Running Title:  A model for trypanosome cell motility




**Abstract:**

The eukaryotic flagellum is a complex and dynamic organelle that provides a cell with highly specialized functions.  The flagellated protozoa are highly dependent on flagellar dynamics for environmental sensing, reproduction, cell morphology, and disease progression. Due to the functional diversity and structural nuances of the flagellum, comparing flagellar motion between different cell types is difficult and a quantitative description of the flagellar dynamics in each flagellated cell type is necessary.  Although some aspects of flagellar motion have been measured in *Chlamydomonas, Crithidia*, *Leishmania, T. cruzi, and T. brucei*, neither a kinematic model of cell motility nor a quantitative description of flagellar dynamics in procyclic *T. brucei* has hitherto been reported.  In this study, we used digital video microscopy to quantify the geometry of trypanosome flagellar waveforms and describe the dynamic properties of the waveforms.  We determined the ranges of amplitude and wavelength in procyclic *T. brucei* and concluded trypanosome flagellar waveforms are dynamic conic functions (ellipses or hyperbolas).  Using the experimental waveform parameters, we derived a novel hyperbolic solution to a nonlinear partial differential equation, which enables us to model trypanosome flagellar motion.  By investigating the types of motion observed in each of four discrete regions of the cell, we propose a step-wise kinematic model of the coordination of flagellar movement and cell motility that results in the auger-like swimming motion of procyclic African trypanosomes. (225)


**Introduction**

The eukaryotic flagellum is a highly conserved yet functionally diverse organelle (for reviews see (Holwill 1974; Dutcher 1995; Cosson 1996; Lindemann and Kanous 1997). The structure (architecture and composition), dynamics (shape changes over time), and regulation (control mechanisms and physical limitations) of the flagellum cooperatively determine flagellar function in each cell. More than a half century of research on the flagellum has revealed that nearly every flagellated cell type exhibits nuances in the structure, use, and / or control of the flagellum. One emerging challenge is identifying the shared characteristics of flagellar motion while elucidating the specific properties of the flagellum unique to each cell type. Undoubtedly, by increasing our understanding of flagellar dynamics, we can advance our ability to elucidate structure-function relationships of molecules within this remarkable organelle.

The movement of flagella and cilia is characterized by a temporally and spatially regulated series of dynamic internally driven bends that result in dynamic waveforms along the filament, and both cilia and flagella can exhibit undulating or oscillatory bends (Cosson 1996). The molecular events that regulate flagellar dynamics are actively being studied, and although several experimentally supported concepts for flagellar regulation exist (Lindemann and Kanous 1997; Brokaw 1999; Dillon and Fauci 2000; Porter and Sale 2000; Cibert 2002; Mitchell 2003), we can not yet speculate on any ubiquitous regulatory mechanisms. Compounding the complexity of flagellar dynamics, flagellar waveform shape is highly variable in different organisms. For example, avian sperm exhibit wide varieties of helical (3D) motion (Vernon and Woolley 1999), whereas mouse sperm flagellar bends are often large amplitude planar propagations with dynamic wavelength and amplitude (Woolley and



Vernon 2002). Though, various models with different mathematical limitations have been employed to describe flagellar dynamics in several cell types, nonlinear behavior is present in flagella (Hines and Blum 1979; Koehler and Powers 2000) and is critical for self-generating swimming patterns via bifurcation or solitary waves propagation along filaments (Kruse et al. 2001; Kruse and Julicher 2003). Similar nonlinear effects are noted in polymers (Sekimoto et al. 1995);(Goldstein and Langer 1995), hair bundles (Boal 2002), or in vitro symmetry breaking mechanisms (Bourdieu et al. 1995) using similar types of nonlinear models. Not surprisingly, three-dimensional flagellar movements also appear to have nonlinearities, as indicated by the observations that a circular helix with linear relationships does not accurately describe the 3D motion of *Xenopus* sperm (Andrietti and Bernardini 1994) and that not all cycles of avian sperm bending lead to production of the same type of 3-D waveform shape (Woolley and Osborn 1984). Nonlinear waveforms can have a variety of shapes and dynamic states that necessitates characterizing the observed parametric limits of the nonlinear forms within an experimental system in order to build equations and expressions to describe that specific system.

In most cell types, the flagellum is designed to move the cell between functionally important environments. Generally, organisms swim in the opposite direction of wave propagation along the flagellum (Holwill 1974), and propulsion velocity of the cell is a fraction of the beat frequency of the flagellum, which is dependent on the amplitude and wavelength of waveform (Cosson et al. 2001). To control swimming direction, the cell must coordinate waveform dynamics with the mechanisms for sensing environmental dynamics (Silflow and Lefebvre 2001). Although techniques are actively being developed to monitor cellular motion in 3 dimensions with as many as four degrees of freedom (Crenshaw 1989;



Blackburn and Fenchel 1999; Crenshaw et al. 2000; Fenchel 2001), high resolution and high speed analysis of cellular motion remains experimentally difficult to quantify. The pinnacle understanding of dynamic cell motility will result from models that co-integrate flagellar dynamics, cellular dynamics, and environmental dynamics.

The trypanosomatids are protozoan parasites that are responsible for several diseases of humans and livestock. Although many environmental, genetic, structural and biochemical differences exist between the various parasites, all trypanosomes are flagellated cells (during at least part of their lifecycle) and dependent on cell motility. *Leishmania sp.* were shown to use their flagella to sense and migrate through osmotic gradients (Leslie et al. 2002), and *T. cruzi* flagellar length was found to be pre-adaptive for differentiation in response to limiting glucose concentrations (Engman et al. 1989). The flagellum of the African trypanosome, *Trypanosoma brucei sp.*, is essential for viability and controls cell morphology, organelle partitioning, cellular differentiation, and attachment to host tissue (Gull 2003). Since cell motility is implicated in disease progression (Hill 2003), the dynamic physical properties of the trypanosome flagellum need to be investigated.

In addition to a conventional 9+2 axoneme, the trypanosome flagellum has several specialized substructures, such as the paraflagellar rod (PFR), flagellar adhesion zone filament (FAZ), and the flagellar pocket, (for review see (Kohl and Gull 1998; Bastin et al. 2000; Gull 2003; Hill 2003; Vaughan and Gull 2003). The integrity of each of the flagellar substructures is critical for productive trypanosome motility (Bastin et al. 1998; Maga and LeBowitz 1999; Maga et al. 1999). For example, *Leishmania sp,.* lacking a functional PFR, exhibit decreased flagellar amplitude, wavelength, and beat frequency resulting in a 4-5-fold reduction in swimming velocity (Santrich et al. 1997), whereas African trypanosomes lacking a PFR are



nearly immotile (Bastin et al. 1998; Bastin et al. 1999). The FAZ in procyclic African trypanosomes is necessary for directional cell motility (Hutchings et al. 2002);(LaCount et al. 2002), and most recently, gamma tubulin was shown to be required for central pair formation and cell motility (Gull 2003; McKean 2003).

Three additional challenges exist when studying cell motility in the African trypanosome. First, the *T. brucei* flagellum is attached to the cell body along its entire length, which restricts flagellar motion and makes flagellar movement codependent on the dynamics of the cell. Second, trypanosomes undergo significant (and in some cases) drastic changes in morphology, behavior, and metabolism during their life cycle. Each of these changes may have implications on the structure, dynamics, and regulation of the flagellum. Third, trypanosomes live in diverse and complex environments ranging from the midgut of an insect to the lymph fluid of a mammal. Each environment has differences in chemical composition and viscosity, which are important factors influencing flagellar dynamics (Hines and Blum 1979);(Walker 1961; Holwill 1974; Woolley and Vernon 2001).

Some properties of flagellar dynamics have been measured in *Crithidia* (Holwill 1974; Holwill and McGregor 1975), *Leishmania* (Santrich et al. 1997)*,* and *T. brucei* (Walker 1961), but a quantitative description of waveform shape and flagellar dynamics in procylic *T. brucei* has not been reported. To understand how African trypanosomes swim in an auger-like fashion, we used digital video microscopy to describe (1) the geometry of trypanosome flagellar waveforms, (2) the dynamic properties of the trypanosome flagellum, and (3) the coordination of flagellar movement and cell motility. The quantification of flagellar waveforms revealed a novel solution to a generalized Korteweg-de Vreis (mKdV) equation which enables modeling of nonlinear flagellar waveforms.



**Materials and Methods:**

*Trypanosome Cell Culture.* *Trypanosoma brucei brucei* strain, Ytat1.1 was maintained at 28° C in SDM-79 medium (JRH Biosciences #60138-1L3965, Lenexa, Kansas) supplemented to 10% with heat inactivated fetal bovine serum. When the culture density was approximately 3-5 x $10^6$ cells/ml, the cells were aliquoted onto the microscope slide directly from the tissue culture flask. When the cell density was in early log phase cells, the cells were concentrated to 1 X $10^7$ cells/ml in 1X phosphate buffered saline (PBS) by centrifugation at 1000 x g for five minutes to increase cell density prior to microscopic investigation. We did not observe any changes in cell behavior or flagellar motion between cells monitored in PBS or growth medium, respectively.

*Video Microscopy.* Microscope slides were cleaned in a chromic acid bath for at least 30 minutes and washed copiously with ultra-pure water prior to cell motility assays. To observe individual flagellar movement, 7-15 microliters of cell suspension was placed under a 22x22 mm cover slip. The smaller the volume, the less freedom of motion we observed in the tethered cells. Cells were visualized using an Olympus BX60 fluorescent microscope equipped for differential interference contrast microscopy. For the light levels and camera sensitivity used in these experiments, the frame rate was generally between 0.25 and 0.4 seconds. Visually healthy cells that were tethered to the glass and appropriately oriented for clear focus on the flagellum were recorded for 20-200 frames. No slide preparation was used for more than 30 minutes to reduce the environmental effects. We intentionally avoided damaged cells and cells lacking significant regions of the flagellum in a single focal plane. Images and digital movies were captured with The Spot™ CCD (Diagnostic Instruments, Inc,



Sterling Heights, MI) and processed using The Spot Advanced (Version 4.0.2.0) and/or Adobe Photoshop 5.5 (Palo Alto, California).

_Analysis of Flagellar Wave Movements._  Approximately 200 digital movies of wild type procyclic form _T. b. brucei_ were collected using differential interference contrast (DIC) microscopy.  Some of the cells were nonspecifically tethered to the glass slides, restricting the movements of the cell and enhancing our ability to resolve discrete flagellar movements. Other cells were recorded while freely swimming to monitor cellular and flagellar movements simultaneously.  Amplitude, wavelength, and perimeter of individual waveforms were measured from isolated frames of digital recordings (figure 1 a and b).  The amplitude of flagellar waveforms was measured as the perpendicular line from the peak of waveform to a straight line that is tangential to the peak of the adjacent waveforms.  The wavelength for each node was measured as the distance between the adjacent nodes (Figure 1).  Translational velocities were measured by dividing the distance the node peak traveled during a series of frames by the elapsed time of the same frames.  Experimental values for flagellar perimeter were measured using the curved line tool in The Spot Advanced software (Version 4.0.2.0). The curve of the flagellum was traced and measured between the points that were used to demark the wavelength.  To calculate the theoretical perimeter (p) of the cell, we used the equation: $p = \pi \dfrac{\sqrt{(a^2 + \lambda^2)}}{2}$, where 'a' is the experimental amplitude of the waveform and 'λ' is the experimental wavelength of the waveform.  Means, standard deviations, and histograms were calculated using Microsoft Excel.

_Mathematical model nonlinear flagellar waveforms_.  Our current model consists of a flexural and torsional thin and finite filament with internal loading and different boundary conditions. The dynamics is described by a generalized Bernoulli-Euler 2-dimensional model of



oscillations of flexural beams (Landau and Lifshiëtis 1967; Magrab 1980). The axial force $P(x,t)$ and transverse force $F(x,t)$ are distributed per unit of length and are dependent on time. This dynamical equation describes the linear approximation of deformation in the filament in space and time, specifically traveling or stationary sine waves. In order to account for the complexity of natural waveform shapes and dynamics, a term was added to account for the local curvature of the filament. We express local curvature as:

$$C \approx y''\left(1 - \frac{3}{2}y'^2 + \frac{15}{8}y'^4 + \ldots\right).$$ In our model the P-force and F-force are proportional to C.

For any given filament, we can use in the following values for the parameters: $J = 1\mu^4$ (cross-section moment of inertia), $E = 100 N/m^4$ (modulus of elasticity), and $m = 1.6 \cdot 10^{-10} Kg/m$ (mass per unit length). The forces can be estimated by multiplying C with a constant, $k \approx -100 pN$. For traveling wave solutions with velocity (V), the equation can be reduced to the following nonlinear and integrable modified Korteweg-de Vries equation (mKdV) (Lamb 1980; Drazin and Johnson 1989; Drazin 1992):

$$\frac{k - P - mV^2}{EJ}\eta' + \frac{3(P+k)}{2EJ}\eta'\eta^2 - \frac{15P}{8EJ}\eta'\eta^4 + \eta''' = 0.$$ Depending on the relative magnitude and sign (±) of the coefficients (i.e. K,P,E,J,V, and m) the solutions of this equation include $sin-$waves, cnoidal waves, isolated solutions with positive amplitude (solitons) or negative amplitude (depression waves or anti-solitons), or combinations of these shapes (Lamb 1980; Drazin and Johnson 1989; Drazin 1992). Considering three or more nonlinear terms in the curvature expression, C, produces novel solutions that include all conic functions, such as ellipses, hyperbola, and parabolas, (figure 5 and Ludu and Hutchings, manuscript in preparation).



## Results:

**Trypanosome flagellar waveform shapes**

Measurements of the amplitude, wavelength, and perimeter of individual waveforms confirm that trypanosome flagellar waveforms do not have a sinusoidal shape (Ludu and Hutchings, manuscript in preparation), but rather are better described by conic functions, such as ellipses and hyperbolas (figure 1). For our analysis, the wavelength ($\lambda$) is equal to the major axis, and a perpendicular line from the apex of the curve to the center of the arc defines the amplitude, 'a' (figure 1). We measured the experimental perimeter of the waveform by tracing the arc-length from the base of the curve to the opposite base of the curve, and we modified the equation for elliptical perimeter (materials and methods) to calculate the theoretical perimeter using the experimental values for the amplitude and wavelength. To determine how well conic functions describe trypanosome flagellar waveform shapes, we compared the theoretical perimeter with our experimentally measured perimeter. The values for the theoretical and the experimental perimeter of a waveform ranged from 87% to 100% accurate; with an average accuracy of $97 \pm 6\%$ (figure 2). These results indicate trypanosome flagellar waveforms can be described as conic functions and that conic equations have predictive power for trypanosome flagellar waveform shape using only two variables.

We observed a Gaussian-like distribution of values for both amplitude and wavelength. The average amplitude in procyclic *T.brucei* cells is $0.742 \pm 0.3$ microns with a range between $0.3 - 2.2$ microns, and the average wavelength is $3.25 \pm 0.95$ microns with a range between 1.24-8.21 microns (figure 3). To determine if flagellar waveforms have a common shape, amplitude and wavelength was plotted for 165 waveforms (figure 3). The shape of a waveform (eccentricity of the arc) can be determined by the ratio of the wavelength



to the amplitude for each waveform.  Therefore, points that cluster along a common line originating from the graph origin share the same relative sized major and minor axes.  African trypanosome flagellar waveforms do not cluster around common shapes, but rather, appear as a continuous range of elliptical shapes, from circular to elongated ellipsoids (figure 3 and 4).

Since flagellar waveform shapes appear to represent a continuous variable within the ranged indicated in figure 3, we graphed the distribution of waveform shapes by the relative abundance of similar shapes (figure 4).  The average eccentricity in our data set is $3.4 \pm 0.77$ microns with a range from $2.12 – 10.1$ microns, yet 55% of the waveform shapes have wavelength to amplitude ratio between 3.0 and 5.0.

**Trypanosome flagellar waveform dynamics**

The dynamic characteristics of individual waveforms were monitored by measuring the amplitude, wavelength and velocity of a waveform in sequential frames of a digital movie. Our results indicate that flagellar waveforms in trypanosomes can have a cohesive shape that translates along the filament but can also have degenerative shape that dilates or constricts as it translates (figure 5 and supplemental movies 1 and 2). An isolated cohesive waveform is known as a soliton or kink and represents one solution to the theoretical model; whereas a degenerative waveform is known as a cnoidal wave (a deformed sine wave), which represents another solution to the same model equation (figure 5).  Using the experimental waveform parameters to solve a novel generalization of the modified Korteweg-de Vreis (mKdV) equation, strongly suggests trypanosome flagellar waveforms consist of nonlinear waves, solitary waves and even combinations of conic functions, like ellipse and hyperbola arcs (figure 5 and Ludu and Hutchings, manuscript in preparation).  Additionally, a single trypanosome flagellum can simultaneously exhibit a complex variety of shapes, including



stationary oscillations and traveling waves. Such waveform shapes can be stable for extended periods but can also suddenly destabilize and change into another pattern (not shown). This suggests that the individual waveforms can exhibit bifurcations, which are another signature of a high degree of nonlinearity. Based on the mKdv equation, consideration of high-order nonlinear terms revealed the existence of piecewise connected conic functions as a real solution to the model (figure 5 and material and methods). Interestingly, hyperbolic functions within these parameters are stable solutions, whereas, elliptical arcs are unstable solutions to the equation (Ludu and Hutchings, manuscript in preparation). For our model to realistically describe trypanosome flagellar waveforms, the scale of the amplitude, wavelength, and forces must coincide with experimental values. Using the results herein, material constants, and force parameters from the literature (Lindemann and Kanous 1997; Boal 2002; Kruse and Julicher 2003), we can conclude that our model produces plausible scale waveform which closely match the experimental observations.

Although most flagellated cells propagate flagellar waveforms from base-to-tip (Lindemann and Kanous 1997), previous reports suggested that blood form *T. brucei* propagates waveforms from tip-to-base and *Crithidia* can propagate waves bidirectionally (Walker 1961; Holwill and McGregor 1975; Maga and LeBowitz 1999). Our results clearly indicate that procyclic African trypanosome flagellar waveforms can travel in both directions along the flagellum (figure 5). Although a free-swimming trypanosome generates an overall impression of a tip-to-base cellular undulation and left-handed helical motion (see figure 6), we have observed individual waveforms along the flagellum propagating in both directions in all regions of the flagellum (figure 5 and data not shown).

**Trypanosome cellular dynamics**



The trypanosome flagellum has a natural left-handed twist and is attached along the entire longitudinal axis of the cell body, and *T. brucei* with unattached flagella are not capable of productive motility (Hutchings et al. 2002; LaCount et al. 2002). Thus in trypanosomes, the geometry and dynamics of the flagellum can not be considered independent of the geometry and dynamics of the cell body. Therefore, an accurate model for trypanosome motility should account for a 2+1 dimensional dynamical equation (coupled surface = 2D + filament = 1D). Toward this end, we examined nearly 200 movies of trypanosome motility and characterized the geometry of the cell relative to the shape of the flagellum. Our results revealed four regions of the cell that exhibit different dynamic characteristics (figure 6).

We defined the 'posterior' region of the cell as the perpendicular cross section from the flagellar pocket to the posterior tip. During periods of cell motility, the posterior region is passive and nearly non-dynamic since there is no part of the flagellum directly associated with this region of the cell body. In free-swimming cells, the posterior end will generally tilt the same direction as the distal end of the flagellum as the anterior end oscillates to conserve the position of the center of mass (figure 6 and supplemental movies 3 and 4).

The 'proximal flagellum' region of the cell extends from the flagellar pocket to the point where the flagellum attaches to a reciprocal position on the opposite side of the cell body (figure 6). Likely due to the mass of the cell body and the spatial restrictions imposed by the flagellar adhesion zone filament, the proximal flagellum exhibits primarily planar oscillations (vibrations) and isolated waveforms. The waveforms in this region can be either symmetric or asymmetric. The anterior boundary of the cell body in this region is often demarked by a slight constriction or bulge where the cell body begins to sharply taper.



Perhaps the most dynamic region of the flagellum is the 'hinge region' (figure 6). The hinge region is a transition zone between the proximal flagellum and anterior flagellum region. Due to the tapered nature of the cell body, the hinge region exhibits a drastic change in curvature and direction relative to the rest of the flagellum. Our observations suggest the hinge region is responsible for controlling the gyre (twist) and angular displacement (crookedness) of the anterior end of the cell.

The 'anterior region' of the cell extends from the hinge region to the distal tip of the flagellum. The dynamic motion of this region contains traveling dynamic waveforms combined with helical rotations or planar oscillations. The movements in the anterior region are more dynamic and more complex relative to the movements observed in the other regions of the cell. The waveforms in this region tend to propagate (predominantly from tip toward the base) but appear to be strongly influenced by the oscillations or gyre localized in the hinge region. The oscillations and rotations in this region are often as high as 45 degrees on either side of the cell's midline and can occur with high frequency (>20 Hz, unpublished observation).

Based on our observations, we have created an idealized animation of the total cellular motion in a typical swimming trypanosome (Figure 6 and supplemental movie 4). We propose that the cell must experience 6 shape transitions to complete one cellular rotation (figure 6). If a resting cell has a natural left-handed twist in the flagellum (figure 6a and d), the first subsequent movement in productive motility is to rotate the anterior region of the cell counterclockwise and generate a super-twist, which increases the gyre of the hinge region (figure 6b and e). The increased gyre in the hinge region induces counterclockwise uncoiling of the posterior region cell, which results in rolling the cell 180 degrees and straightening the



flagellum (figure 6c and f).  Next by rotating the anterior region upward, the flagellum will reassume its natural half-helical twist.  Upon super-twisting the hinge region, the cell will uncoil and roll 180 degrees counterclockwise resulting in a straight flagellum (figure 6h).  The cell will assume its natural left-handed helical shape and has completed one cellular rotation.



**Discussion:**

Using a geometric approach to describe flagellar waveforms has proven to be a successful means to reduce the complexity of the whole flagellum into its definable parametric attributes since Brokaw fit complex flagellar shapes to a series of circular arcs connected by straight lines (Brokaw 1965; Eshel and Brokaw 1988). Subsequently, a geometric approach was used to describe the nature of the waveform shapes in the trypanosomatid, *Crithidia* (Holwill and McGregor 1975), and a three-dimensional application of arcs and lines was used to describe the helical shape of a flagellum (Crenshaw 1989). Most recently, the planar waveforms of sea urchin sperm were modeled as a series of circular tangential waveforms (Cibert 2002). Thus, describing the geometry of the flagellum is an effective means to deduce flagellar dynamics at the cellular level without having the molecular resolution of specific molecular structures and regulatory complexes. In future studies, we can test, compare, and integrate the conic function descriptions of trypanosome waveforms with the geometric arguments being established for other organisms.

Although quantitative analysis of waveforms has been conducted on the biflagellated green algae, *Chlamydomonas rheinhardti* (Brokaw and Luck 1983; Omoto and Brokaw 1985), of the trypanosomatids, quantitative analysis has been conducted only in *Crithidia* (Holwill 1974). However, several groups have described general changes in beat frequencies and swimming velocities in *T.brucei, T. cruzi,* and *Leishmania* (Maga and LeBowitz 1999; Bastin et al. 2000; Hill 2003). The average amplitude and range of values that we observed for flagellar waveforms in African trypanosomes are significantly different than the amplitudes and wavelengths previously reported for *Crithidia oncopelti* and *Leishmania*. In *C. oncopelti,* the average amplitude and wavelength are 2.4 microns and 14.4 microns,



respectively (Holwill and Silvester 1965), and the reported wavelengths in *Leishmania* are greater than 10 microns (Holwill and Silvester 1965; Santrich et al. 1997). We determined the average trypanosome amplitude and wavelength of 0.74 and 3.25 microns, respectively. The apparent discrepancy in values may be due to differences in flagellar attachment, flagellar dynamics, and / or flagellar regulation. A side by side experimental comparison of flagellar waveforms and cell motility may enable identification of the primary differences in flagellar dynamics between these related organisms.

*C. oncopelti* normally produce planar waves that propagate from tip to base and increase in wavelength and amplitude as they propagate (Holwill and Silvester 1965), and the same has been concluded for African trypanosomes (Bastin et al. 2000; Hill 2003). Our data suggests that procyclic African trypanosome flagella can generate waves in both directions and the wavelength and amplitude do not significantly increase during propagation. Furthermore, each of the various waveform shapes are not equally abundant (figure 4), which suggests that some restrictions are imposed on the proportionality of the amplitude and wavelength (eccentricity). A waveform with a large amplitude or large wavelength is not predisposed to have a very high or very low eccentricity. Some of the most atypical shapes were found in waveforms of near average wavelength, but with disproportionately small or large amplitude, respectively. Likewise, some of the largest wavelengths were associated with waveforms with average eccentricity. We anticipate that waveform shape will influence cell motility, but our current analysis does not afford us to determine the influence of waveform shape on overall cell motility

Many of the cells that we examined were tethered to the glass slide, restricting their ability to swim out of focus. In *C. oncopelti* cells that were attached to glass exhibited changes in the



symmetry of the waves, but the flagellum exhibited a normal frequency, amplitude, and wavelength (Holwill and Silvester 1965). In blood form *T. brucei gambiense* and *T. brucei rhodesiense* that were attached to glass, waves either propagated through an immobilized region of the flagellum or initiated intrinsically within the flagellum on both sides of an immobilized region (Walker 1961). Our results confirm this observation in procyclic *T. brucei*, suggesting the underlying mechanism for flagellar bend formation in trypanosomes may be independent of flagellar geometry. Several groups are currently investigating the signals that underlie bend formation and waveform propagation in various cell types (Hines and Blum 1983; Julicher and Prost 1995; Lindemann and Kanous 1997; Omoto et al. 1999; Mitchell 2003; Wargo and Smith 2003).

During migration from the midgut to the salivary glands of the tsetse, the parasite must use the flagellum to attach to host tissue, penetrate host tissue, and migrate through chemically dynamic environments (Hill 2003). Therefore, investigating flagellar dynamics in tethered cells may shed light on flagellar dynamics that occur during transitional phases of the parasite's lifecycle. Furthermore, waveforms traveling through the hinge region of the cell can either generate helical (3D) twist or planar oscillation (figure 6). Because trypanosomes experience chemically and viscously dynamic environments throughout their lifecycle, both planar and helical modes of motility may be necessary for these cells.

A quantitative description of flagellar motion in wild type cells enables the investigation of the changes in motility that occur in different conditions, such as motility mutants (ie. flagellar adhesion zone and paraflagellar rod mutants: Hutchings et al. 2002; LaCount et al. 2002; McKean et al. 2003) and lifecycle stages. Additionally, a mathematical model of flagellar dynamics increases our resolution of the characteristics that are responsible



for the changes in flagellar motion. Our model allows us to treat each waveform as an isolated conic function, but also enables us to consider the overall shape, velocity, and relative position of a set of coexisting waveforms as they cooperatively influence the global dynamics of the flagellum. Thus, our model can accommodate complex shapes along the total filament. These shapes can be modeled through the inclusion of additional nonlinear terms in the mKdV equation. However, as additional nonlinear terms are considered in the model, the stability of the waveforms decreases (not shown). An experimental result that coincides with the theoretical argument, such as that described herein, enhances the overall validity of the model. Thus, a 'good' model balances the complexity and stability, resulting in solutions that are reasonable for a natural (stable) system.

In summary, digital video recordings of wild type procyclic *T. brucei* were analyzed to describe flagellar waveform shape, dynamics, and cell motility. We have provided a quantitative standard for flagellar waveform shapes in procyclic *T. brucei*, and we have shown that trypanosome flagellar waveforms are accurately described by stable or unstable translating conic functions. To enable mathematical simulation of trypanosome waveforms, we have derived a novel set of solutions to the nonlinear differential equation that governs flagellar dynamics. The solutions include traditional sine (linear approximation), soliton, and anti-soliton (mKdV) waves and also include a new solution that is characterized by a tangentially connected series of conic arcs (highly nonlinear). Based on our observations of motile cells, we have proposed the first kinematic model for the coordination of flagellar dynamics and cell shape to describe the natural auger-like swimming motion of procyclic African trypanosomes.



**Acknowledgements:**


We greatly appreciate the input and contributions of Dr. Darrell Fry, IDEAS program Co-director, http://scitech.nsula.edu/ideas.  We are grateful to Northwestern State University, The College of Science and Technology, The Departments of Biological Sciences and Chemistry and Physics for funding assistance.  We would like to especially thank our student research assistants: Anna Jones, Jeffrey Cain, Angela Rivers, Josh Sonnier, Courtney Hebert, Lori Langlinais, Justin Raines, Lydia Archuletta, Ashley Dunham, and Caryn Ratcliff for their contributions to this work.  This work was partially funded by the Louisiana Board of Regents Research Competitiveness Subprogram Proposal: LEQSF(2004-2007)-RD-A-25, to Drs Hutchings and Ludu, and the NSU Research Council.  Dr. Ludu is also supported by NSF grant: NSF-0140274.

**Figure 1.  Trypanosome flagellar waveforms and the properties of an ellipse**
(1A) Differential interference contrast (DIC) micrograph of a growing procyclic form

*Trypanosoma brucei.*  The white box in indicates an isolated waveform.  (1B) Magnified

image of the example trypanosome flagellar waveform outlined in 1A.  To understand the

geometry of flagellar waveforms, the amplitude (a), wavelength ($\lambda$), and perimeter (p) were

measured as described.  (1C) Diagram of the properties of an ellipse, which is a good fit to

trypanosome flagellar waveform shape.  An ellipse differs from a circle in that an ellipse has

two foci ($F_1$ and $F_2$), which determine the perimeter (p) according to the length of the two

radii ($r_1$ and $r_2$) extending from each respective focus.  The sum of the radii ($r_1+r_2$) is equal to

the wavelength ($\lambda$, major axis), and when $r_1$ and $r_2$ are equal length, a right triangle is formed

between the focus, the centroid (gray circle), and the intersection of the radii on the perimeter.

Using the Pythagorean Theorem, one can solve for (b), which is the distance from the centroid

to the focus.  We can define the distance between foci as segment c, which equals 2b.  By

measuring only two variables, the amplitude and wavelength, one can calculate all the

properties of the ellipse.

**Figure 2.  Perimeter of an ellipse describes flagellar waveform shape**

The amplitude and wavelength of individual waveforms accurately predicts the perimeter of

the waveform shape using a modified equation for the perimeter of an ellipse. The theoretical

perimeter (black bars) and the experimentally measured perimeter (gray bars) of 30 example

waveforms are shown. The values of the theoretical and experimental perimeters of each

waveform range from 87% to 100% accurate, with an average accuracy of 97%.  For each

waveform, the theoretical perimeter was predicted from the values for amplitude and



wavelength using the equations described in figure 1. The experimental perimeter was measured from DIC micrographs of individual cells.

**Figure 3. Distribution of flagellar waveform shapes found in trypanosomes**

Amplitude and wavelength of 164 flagellar waveforms are plotted. The gray bar on the x and y-axes represents the range of wavelength and amplitude, respectively, and the gray arrow indicates the average for each parameter. In *T. brucei,* the average amplitude is 0.742 +/- 0.3 microns with a range of 0.3 – 2.2 microns, and the average wavelength is 3.25 +/- 0.95 microns with a range of 1.24-8.21 microns. The shape of an ellipse (eccentricity) is determined by the proportionality of the amplitude and the wavelength. All points on the graph that are intersected by a straight line (common slope) will have the same elliptical shape. The dashed lines represent ellipses with wavelength : amplitude ratios of 2.0 (circular) to 10.0 (elongated ellipsoid) for reference. The amplitudes and wavelengths were measured from 100X DIC micrographs of log phase procyclic *T. b. brucei* as described in the materials and methods.

**Figure 4. Distribution of flagellar waveform shapes**

More than 50% of the flagellar waveforms measured have a wavelength to amplitude ratio between 3.5 and 5.0, although the eccentricities of flagellar waveforms range from near circular (2.0) to dilated ellipses (10.0). The histogram shows the distribution of flagellar waveforms within the indicated ranges. The average eccentricity is 2.4 +/- 0.77 with a range from 2.12 – 10.1. The diagram at the top shows the shape of ellipses within the observed



ranges. Dashed lines connect each column in the histogram to the respective shape in the diagram.

## Figure 5.  Trypanosome flagellar waveforms are dynamic ellipses

Trypanosome flagellar waveforms can be described as translating, dilating, and/or constricting conic functions.  Panels A-C show an example of a translating elliptical waveform. The white arrow indicated the location of the waveform in each frame, and the white ellipse is a scaled representation of the ellipse predicted by the waveform. In this movie, the waveform is translating from base-to-tip at approximately 1.5 microns per second.  Panels D-F show an example of a dilating/constricting waveform. The white arrow waveform is constricting (decreasing wavelength over time) and the black arrow waveform is dilating (increasing wavelength over time). The gray and black ellipses in each frame represent the shape of the respective waveforms. At the bottom of each frame, the elapsed time is indicated (hrs: min: sec. millisec). To view the flagellar motion in these cells, please view *supplemental movies 1 and 2*.  Panel G shows four possible solutions of the generalized mKdV equation that describes flagellar waveforms.  (a) Represents a linear approximation of a sine wave.  (b) Shows a cnoidal wave, which is a deformed sine wave.  (c) Represents a soliton, a nonlinear isolated cohesive traveling waveform, and (d) Shows a kink, which is a second type of soliton solution.  Panel H represents a novel solution to the generalized mKdV equation based on our experimental observations.  The waveform is a series of tangentially connected hyperbolic arcs.

## Figure 6. A model for the auger-like swimming motion of *T. brucei*



The auger-like swimming motion of *T. brucei* can be described by a twist, super twist, and roll model. Panels A and C depict the natural shape of trypanosome, which is a left-handed half helix. Panels B and E depict a cell that has super twisted the flagellum into a full helix, which will subsequently cause rotation (uncoiling) of the cell body. Panels C and F depict a cell that has been uncoiled, and the flagellum is straightened along the longitudinal axis of the cell. Panel G is a diagram of a procyclic form trypanosome indicating the features in the cell body that demark segmentation of the cell based on motility characteristics. The 'posterior' region of the cell is demarked from the flagellar pocket to the posterior tip. The 'proximal flagellum' region of the cell is between the flagellar pocket and the point on the cell body where the flagellum contacts a reciprocal position on the opposite side. The anterior boundary of this region is often demarked by a slight curve or constriction in the cell body shape. The 'hinge' region is between the posterior flagellum and anterior flagellum, and is demarked by a dramatic change in curvature of the flagellum. The hinge region of the cell exhibits dynamic oscillations and helical twists. Panel H shows an example composite of the six cell orientations that can produce auger-like motion. For an animated version of the model, please download *supplemental movie 3 and 4.*





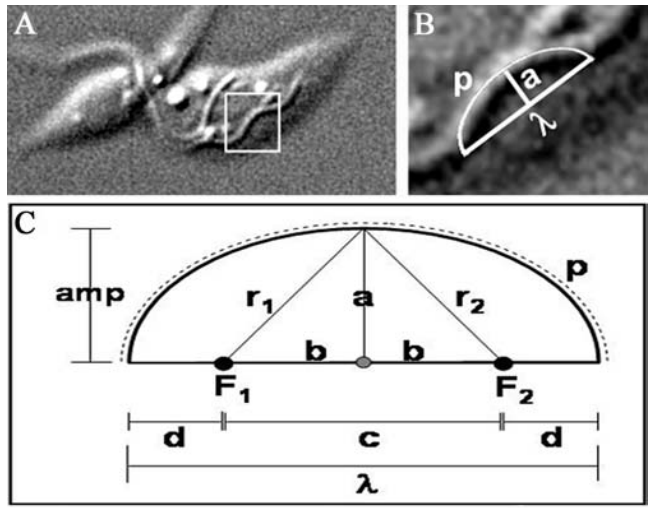



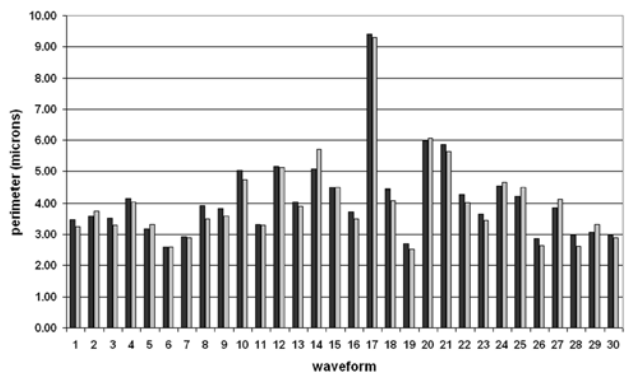



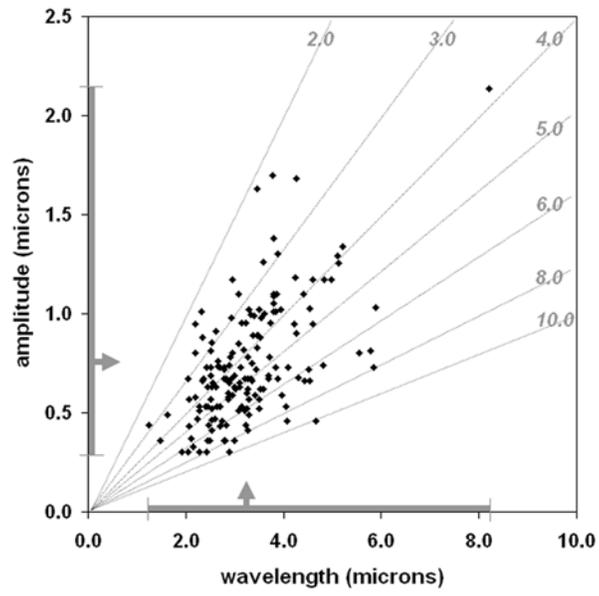



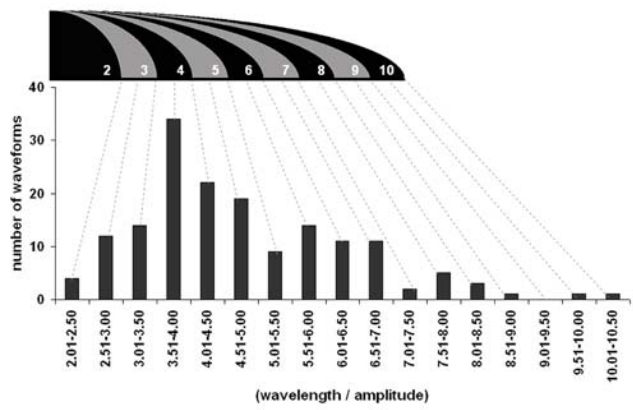



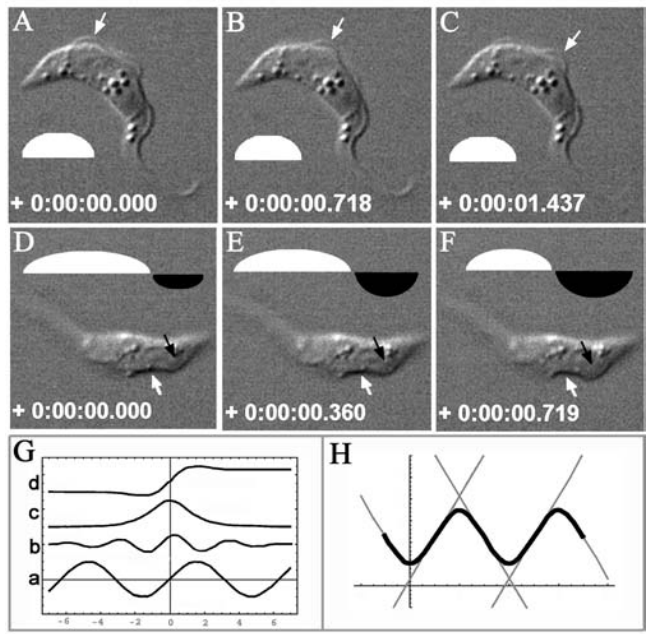



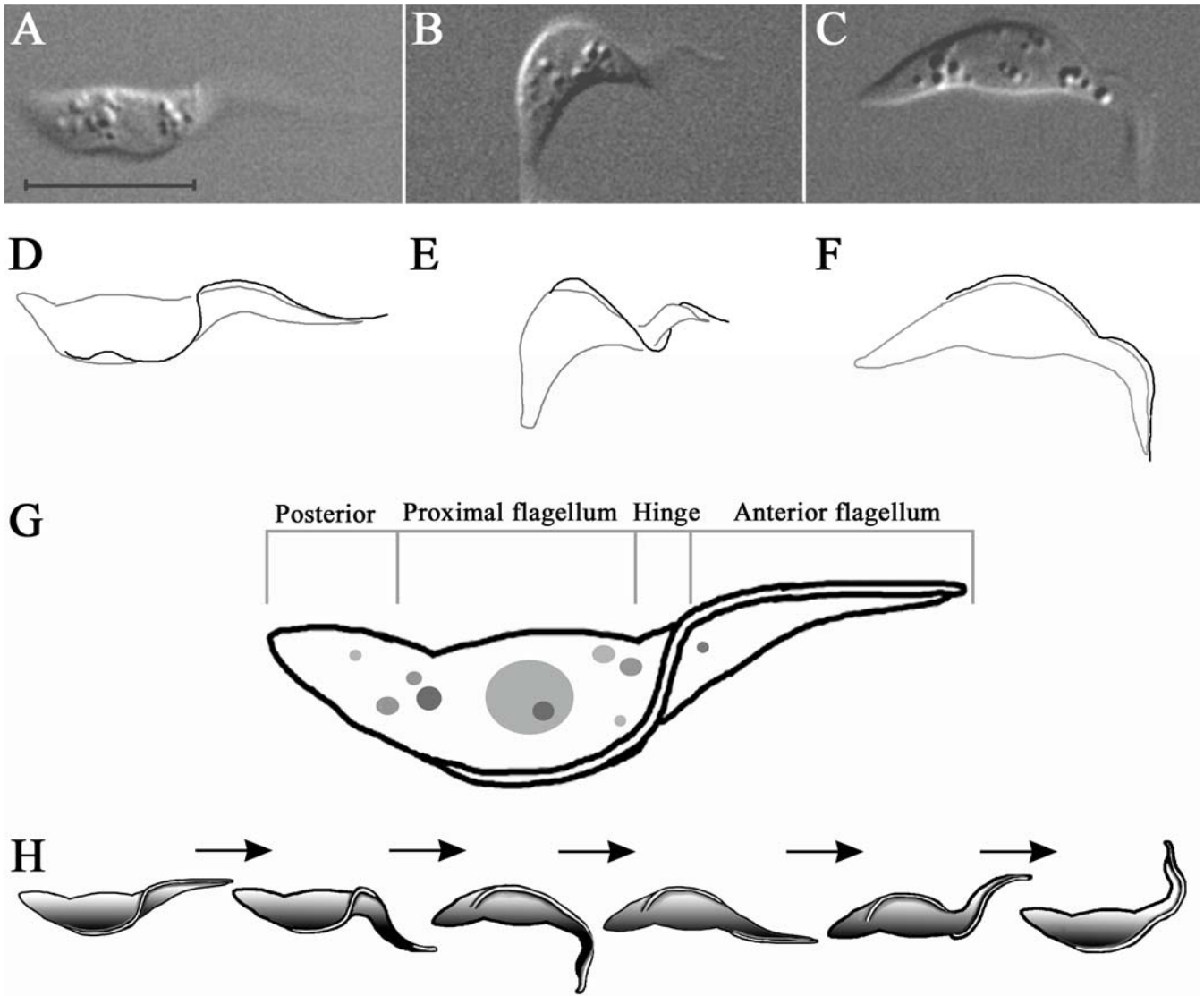

G

Posterior | Proximal flagellum | Hinge | Anterior flagellum